\documentclass[aps,prl,reprint,notitlepage,floatfix,superscriptaddress]{revtex4-2}
\usepackage{amsmath,amssymb,bm,graphicx,booktabs,placeins,needspace,hyperref,microtype}
\hypersetup{hidelinks,pdfauthor={Xing-Hui Feng and Jun Peng},pdftitle={What Schwarzschild Ringdown Can and Cannot Reveal about Gravity}}
\newcommand{\dd}{\mathrm d}
\newcommand{\eps}{\epsilon}
\newcommand{\Tr}{\operatorname{tr}}
\newcommand{\Ree}{\operatorname{Re}}
\newcommand{\Imm}{\operatorname{Im}}
\begin{document}
\title{What Schwarzschild Ringdown Can and Cannot Reveal about Gravity}
\author{Xing-Hui Feng}
\email{xhfeng@tju.edu.cn}
\affiliation{Center for Joint Quantum Studies and Department of Physics, School of Science, Tianjin University, Tianjin 300350, China}
\author{Jun Peng}
\email{junpeng@bjut.edu.cn}
\affiliation{School of Physics and Optoelectronic Engineering, Beijing University of Technology, Beijing 100124, China}
\date{November 29, 2026}
\begin{abstract}
Even a complete quasinormal-mode (QNM) spectrum need not determine the gravitational action. For parity-invariant algebraic curvature interactions in four-dimensional vacuum gravity, the leading and subleading real-frequency responses in the first-order eikonal expansion distinguish all independent Schwarzschild curvature responses at each degree above two, at fixed ADM mass and overtone. At degree six, a numerical calculation shows that the two quadrupolar fundamental modes already determine these responses. Nevertheless, a sixth-degree interaction leaves the linearized vacuum equations unchanged while generating a nonzero on-shell mixed-helicity six-graviton amplitude. Thus even an exact linear spectrum cannot recover all physical information in the action.
\end{abstract}
\maketitle
\clubpenalty=10000
\widowpenalty=10000
\displaywidowpenalty=10000

Black-hole spectroscopy tests general relativity (GR) through the frequencies and damping rates of quasinormal modes (QNMs) \cite{Dreyer,BertiWill,GW250114}. In GR the odd- and even-parity equations of Schwarzschild have different potentials but the same QNM spectrum \cite{ReggeWheeler,Zerilli,Chandrasekhar75,MP}. Some curvature terms do not enter the spherical field equations \cite{Chen}; those in Ref.~\cite{Nakashi}, for instance, preserve the Schwarzschild metric but break isospectrality. Even if the mass and all the QNM frequencies were known, which interactions could be distinguished?

Parametrized ringdown relates changes in the master potentials to frequency shifts \cite{ParamRingdown}. Perturbation equations are known for cubic and quartic interactions \cite{deRham,Cardoso,CanoSlow,CanoWave}, and general Teukolsky formalisms are available for rotating black holes \cite{CanoTeuk,LiTeuk}. Rather than computing the spectrum for a specified gravitational action, we ask how much information about the action can be recovered from the spectrum itself. At each curvature degree a static spherical metric samples only one combination of Weyl invariants \cite{Bueno}; linear perturbations can probe additional combinations. Different perturbation equations may yield identical QNM frequencies, and physically inequivalent curvature interactions may lead to the same linearized field equations on Schwarzschild.

\textit{Curvature responses.---}Consider the vacuum action
\begin{equation}
 S=\frac{1}{16\pi G}\int\dd^4x\sqrt{-g}
 [R+\epsilon\mathcal F(R_{abcd})].\label{action}
\end{equation}
The polynomial $\mathcal F$ is invariant under spatial parity and includes the higher-curvature coupling constants; $\epsilon$ counts EFT order \cite{Donoghue}. We exclude independent curvature derivatives and nonlocal terms. On the asymptotically flat GR-connected branch, the ADM mass $M$ and asymptotic time normalization are fixed. Local perturbative field redefinitions remove Ricci-dependent operators in vacuum \cite{Ruhdorfer,CriadoPV}. At finite coupling, curvatures and perturbation momenta must remain within the EFT regime.

In the Weyl representative, set $y=r/M$, measure curvature in units of $M^{-2}$, and rescale the degree-$N$ couplings by $M^{-(2N-2)}$ in $F$. In an oriented orthonormal frame define the electric and magnetic Weyl tensors by $E_{ij}=C_{\hat0 i\hat0 j}$ and $B_{ij}=\varepsilon_i{}^{kl}C_{klj\hat0}/2$. We use the invariants
\begin{equation}
 \begin{aligned}
 \mathsf I&=\Tr[(E+iB)^2]=X+iP,\\
 \mathsf J&=\Tr[(E+iB)^3]=Y+iQ.
 \end{aligned}
 \label{invariants}
\end{equation}
The invariants $X,Y,P,Q$ generate the Weyl polynomials \cite{Ruhdorfer}. On Schwarzschild, $E+iB=u\,\mathrm{diag}(-2,1,1)$, with $u=y^{-3}$, so $(X,Y,P,Q)=(6u^2,-6u^3,0,0)$. The value of the interaction $F(X,Y,P,Q)$ and its curvature expansion through quadratic order are determined by three functions:
\begin{equation}
 \begin{aligned}
 F_0(u)&=F(6u^2,-6u^3,0,0),\\
 \tau(u)&=9uF_Y\big|_0,\\
 \mu(u)&=\tau+\big[72u^2F_{PP}-216u^3F_{PQ}+162u^4F_{QQ}\big]_0 .
 \end{aligned}
 \label{responses}
\end{equation}
The subscript $0$ denotes evaluation at the Schwarzschild values of the Weyl invariants (see Eq.~(II.10) of Ref.~\cite{companion} for the expansion). In this Weyl frame, $F_0$ determines the correction to the static spherical background and contributes to both parities. The magnetic response $\mu$ couples to radial magnetic curvature and affects only odd parity; the radiative response $\tau$ couples the extreme Weyl scalars of the Newman--Penrose formalism \cite{NewmanPenrose}. When varied independently of $F_0$, these responses change the perturbation equations without affecting the background metric.

Quadratic-curvature terms are redundant in vacuum EFT modulo field redefinitions and the Gauss--Bonnet term. The extra massive modes of the full theory \cite{AntoniouQuadratic,AntoniouSuppression} lie outside our perturbative branch. For a homogeneous interaction of degree $N>2$, the responses take the form $(c_0u^N,c_\mu u^{N-2},c_\tau u^{N-2})$. The number of independent curvature responses depends on the curvature degree:
\begin{equation}
 r_N=\begin{cases}
 1,& N=3,\\
 2,& N=4,5,7,\\
 3,& N=6\ \text{or}\ N\geq8.
 \end{cases}\label{rank}
\end{equation}
At degree six, $X^3$, $XP^2$ and the discriminant $\mathcal D_6=\Ree(\mathsf I^3-6\mathsf J^2)$ \cite{Baker} give pure background, magnetic and radiative responses, respectively. This is the first degree at which a nonzero curvature polynomial can have all three responses vanish. At degree seven, $2XPQ-3YP^2$ is another interaction with vanishing responses. For the derivation of Eq.~\eqref{rank} and the full classification, see Appendices A.1 and A.2 of Ref.~\cite{companion}, respectively.

\textit{Curvature information in the QNM spectrum.---}At fixed ADM mass and curvature degree, interactions with the same three curvature responses give identical first-order QNM shifts. Whether these shifts distinguish all the responses is less obvious, since distinct master potentials may be isospectral \cite{LenziDarboux}. We address this question using the order-reduced equations on the GR branch \cite{ParkerSimon,BuenoStability}, derived in Ref.~\cite{companion}. With $\hat\omega=M\omega$, $f=1-2/y$ and $\widehat V=M^2V$, these are
\begin{equation}
 \frac{\dd^2\Psi_\pm}{\dd x^2}+(\hat\omega^2-\widehat V_\pm)\Psi_\pm=0,\quad \frac{\dd x}{\dd y}=(AB)^{-1/2},\label{master}
\end{equation}
The metric functions $A,B$ are defined by $\dd s^2/M^2=-A\,\dd t^2+B^{-1}\dd y^2+y^2\dd\Omega^2$, where $t,x$ are dimensionless and $g=\sqrt{AB}$. With $J_l=l(l+1)$, the coefficient of $\epsilon$ in $\widehat V_\pm$ is
\begin{equation}
 \begin{aligned}
 \Delta\widehat V_-&=\mathcal B_-[F_0]-\frac{fJ_l(J_l-2)}{8y^4}\mu+\mathcal T_-[\tau],\\
 \Delta\widehat V_+&=\mathcal B_+[F_0]+\mathcal T_+[\tau].
 \end{aligned}
 \label{potentials}
\end{equation}
Here $l\geq2$; $\mathcal B_\pm$ includes the metric correction and direct curvature terms, and $\mathcal T_\pm$ describes the radiative coupling. The full radial coefficients are given in Sec.~V and Appendix E of Ref.~\cite{companion}. Its Sec.~VI proves equivalence to the metric problem on the GR-connected low-energy branch for ordinary isolated modes satisfying the reconstruction conditions, away from exceptional frequencies and the zero-frequency cut. The QNM conditions are $\Psi_\pm\sim e^{-i\hat\omega x}$ at the future horizon and $\Psi_\pm\sim e^{i\hat\omega x}$ at infinity, with time dependence $e^{-i\hat\omega t}$.

For the homogeneous interaction set $m=3(N-2)$ and $\phi=F_{0,u}/(96u)$, so that $(\phi,\mu,\tau)=(p,c_\mu,c_\tau)y^{-m}$, with $p=Nc_0/96$. The GR barrier at large $l$ is $J_lv_0$, with $v_0=f/y^2$. The coefficient of $J_l^2$ in the first-order potential is (Sec.~\ref{app:large} of the Supplemental Material~\cite{supplement})
\begin{equation}
 W_- =\frac{f(c_\tau-c_\mu)}{8y^{m+4}},\qquad
 W_+ =-\frac{f(c_\tau+16mp)}{8y^{m+4}}.
 \label{eq:radial_leading}
\end{equation}
The large-multipole behavior need not be fixed by the background null geodesics. At a fixed overtone, we use the first-order WKB expansion at zero coupling before taking the large-$l$ limit. The leading real-frequency responses are
\begin{equation}
 \mathcal E_\pm=\lim_{J_l\to\infty}\frac{1}{J_l}\left.\partial_\epsilon\ln\Ree\hat\omega_\pm\right|_0.
\end{equation}
In GR the leading real frequency is $\Ree\hat\omega_0\sim\sqrt{J_l/27}$, determined by the light ring \cite{CardosoGeodesics}. Higher-curvature interactions modify the graviton dispersion \cite{CanoDavid,CanoEikonal2026}. The barrier-top WKB expansion in Supplemental Sec.~\ref{app:large} gives $\mathcal E_\pm=W_\pm(3)/(2v_0(3))$, hence
\begin{equation}
 \mathcal E_-=\frac{c_\tau-c_\mu}{16\,3^{m+2}},\qquad
 \mathcal E_+=-\frac{c_\tau+16mp}{16\,3^{m+2}}.
 \label{eq:spectral_leading}
\end{equation}
The leading real-frequency shifts coincide for $c_\mu=2c_\tau+16mp$, which includes the equal-coefficient quartic case \cite{CanoDavid} but does not imply finite-$l$ isospectrality. When three responses are independent, the leading frequency shifts of the two parities cannot determine them separately. Both leading shifts vanish when $c_\mu=c_\tau=-16mp$. On this direction, the $J_l^2$ corrections vanish throughout the exterior, and the subleading odd-parity response (Supplemental Eq.~\eqref{nullresponse}) is
\begin{equation}
 \begin{aligned}
 &\lim_{J_l\to\infty}\left.\partial_\epsilon\ln\Ree\hat\omega_-\right|_{\epsilon=0}=\frac{p}{27\,3^m}\\
 &\quad\times\left[\frac{4(N-2)(3+1/N)}{N-1}+2m(m+1)(m+2)\right].
 \end{aligned}
 \label{null}
\end{equation}
The bracket is strictly positive for $N>2$. Thus vanishing spectral responses in the eikonal expansion require $p=c_\mu=c_\tau=0$. Let $\mathcal R_N$ map a degree-$N$ interaction to its response amplitudes and $\mathcal S_N$ to its first-order spectral responses in the WKB expansion in both parities. We obtain
\begin{equation}
 \boxed{\ker\mathcal S_N=\ker\mathcal R_N,\qquad N>2.}\label{spectral_complete}
\end{equation}
At fixed overtone, the leading and subleading real-frequency responses thus distinguish all independent curvature responses at each curvature degree. The large-$l$ limit is taken at first EFT order, not at finite coupling.

\textit{Quadrupolar spectroscopy.---}At degree six, the numerical quadrupolar response provides an independent test. We use the basis $(\phi,\mu,\tau)=(\lambda_0,\lambda_\mu,\lambda_\tau)(2/y)^{12}$, absorbing the small EFT factors into the real $\lambda_i$. The oscillation frequencies and damping rates of the two modes provide four real spectral quantities. Table~\ref{tab:qnm} gives the derivatives from horizon-adapted Chebyshev collocation \cite{Jansen}, checked by Legendre collocation (Supplemental Eq.~\eqref{sm:eigenvalue_derivative}).
\begin{table}[t]
\caption{Dimensionless fundamental quadrupolar responses $\partial_{\lambda_i}(M\omega)$ at degree six, at zero coupling and fixed ADM mass. The basis is $(\phi,\mu,\tau)=(\lambda_0,\lambda_\mu,\lambda_\tau)(2/y)^{12}$, with the small coupling factors absorbed into $\lambda_i$.}
\label{tab:qnm}
\begin{ruledtabular}
\renewcommand{\arraystretch}{1.16}
\begin{tabular}{lrr}
 & \multicolumn{1}{c}{Odd} & \multicolumn{1}{c}{Even}\\\hline
$\lambda_0$ & $-0.08137-0.22228i$ & $0.12756-0.13713i$\\
$\lambda_\mu$ & $0.00046-0.00163i$ & $0$\\
$\lambda_\tau$ & $-0.02138-0.01822i$ & $0.02432+0.00853i$
\end{tabular}
\end{ruledtabular}
\end{table}
With $\delta\hat\omega_\pm=\sum_iR_{\pm i}\lambda_i$ and $R_{\pm i}=\partial_{\lambda_i}\hat\omega_\pm|_0$, the even frequency fixes $(\lambda_0,\lambda_\tau)$: $R_{+\mu}=0$ and its two-dimensional real response matrix has determinant $4.42\times10^{-3}$. The remaining response is $\lambda_\mu=(\delta\hat\omega_--R_{-0}\lambda_0-R_{-\tau}\lambda_\tau)/R_{-\mu}$, with $R_{-\mu}\ne0$. Requiring real $\lambda_\mu$ gives one linear consistency condition on the four real shifts.

\Needspace{6\baselineskip}
The amplitudes are uniquely determined at fixed mass, but frequency errors may be amplified (singular values $(0.302928,0.028158,0.001280)$ in this basis). In particular, the odd mode is much less sensitive to $\lambda_\mu$ than to the other two amplitudes. A measurement also depends on mode excitation and frequency resolution in both parities \cite{LeaverResponse,SilvaExcitation}, as well as correlations with the remnant parameters \cite{MaenautRingdown}.

\begin{figure}[t]
 \centering
 \includegraphics[width=0.97\columnwidth]{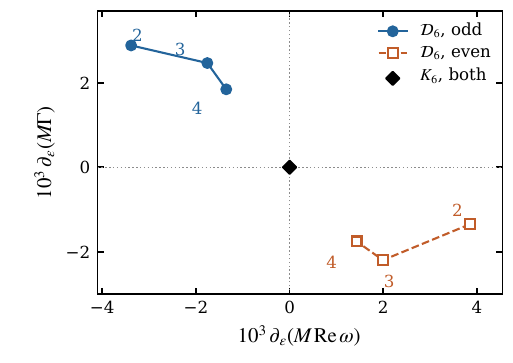}
 \caption{Same geometry, different ringdown. The interactions $\mathcal D_6=\Ree(\mathsf I^3-6\mathsf J^2)$ and $K_6=8Q^2-3XP^2$ both preserve Schwarzschild. Fundamental-mode derivatives for $R+\alpha\mathcal D_6$ and $R+\alpha K_6$ at $\varepsilon=\alpha/M^{10}=0$ are shown for $l=2,3,4$; $\Gamma=-\Imm\omega$. All $K_6$ shifts vanish. For $\mathcal D_6$, $\lambda_\tau=648\varepsilon/2^{12}$.}
 \label{fig:comparison}
\end{figure}

\textit{Interactions invisible to spectroscopy.---}To see how the same geometry can yield different linear dynamics, compare the discriminant $\mathcal D_6$ with
\begin{equation}
 K_6=8Q^2-3XP^2.
 \label{examples}
\end{equation}
Both invariants and their first variations vanish on Schwarzschild, so in this Weyl representative the metric solves either vacuum theory exactly. Their quadratic expansions are different: $\mathcal D_6$ has $F_0=\mu=0$, $\tau=648u^4$, while all three responses vanish for $K_6$. In fact, the background Weyl matrix satisfies $\mathsf W_0^2=2u^2\mathbf 1-u\mathsf W_0$, where $\mathsf W=E+iB$. Using $\Tr\delta\mathsf W=0$ in the first variations of Eq.~\eqref{invariants} gives $\delta\mathsf J=-3u\delta\mathsf I/2$, hence $\delta Q=-3u\delta P/2$ and $[K_6]_2=8(\delta Q)^2-3\bar X(\delta P)^2=0$ at every radius. Figure~\ref{fig:comparison} illustrates the parity splitting due to $\mathcal D_6$, despite the unchanged photon sphere at $r=3M$. For $K_6$ the linearized vacuum equations themselves remain those of GR.

The vanishing quadratic response does not imply that $K_6$ is redundant under field redefinitions. In terms of the self-dual and anti-self-dual invariants,
\begin{equation}
 K_6=-2(\mathsf J-\overline{\mathsf J})^2
 +\frac38(\mathsf I+\overline{\mathsf I})
 (\mathsf I-\overline{\mathsf I})^2 .\label{chiral}
\end{equation}
The term $4\mathsf J\overline{\mathsf J}$ contributes at six points with three helicities of each sign. A linearized chiral Weyl spinor is proportional to $\ell_A\ell_B\ell_C\ell_D$; each pair in the cyclic cubic contraction supplies two spinor brackets \cite{Ruhdorfer}. With the conjugate chirality for legs 4--6 this gives
\begin{equation}
 \mathcal A^{K_6}_{---+++}\propto
 \langle12\rangle^2\langle23\rangle^2\langle31\rangle^2
 [45]^2[56]^2[64]^2.\label{amplitude}
\end{equation}
The other chiral terms have even helicity multiplicities and cannot cancel this contribution.

For complex null momenta $k_i=\ell_i^{\mathsf T}\tilde\ell_i$, take the row spinors in order $i=1,\ldots,6$:
\begin{equation*}
\begin{aligned}
(\ell_i)&=((1,0),(0,1),(1,1),(1,2),(2,1),(3,1)),\\
(\tilde\ell_i)&=((-6,-8),(-5,-5),(2,3),(1,0),(0,1),(1,1)).
\end{aligned}
\end{equation*}
Here $\sum_i k_i=0$, and the bracket product in Eq.~\eqref{amplitude} equals one. Since $K_6$ begins at sixth order in metric perturbations about flat spacetime, its leading tree-level correction to six-graviton scattering is a contact interaction. The nonzero on-shell amplitude shows that $K_6$ cannot be eliminated by a local perturbative field redefinition or a total derivative \cite{CriadoPV}.

\textit{Discussion.---}In the chosen Weyl representative, the Schwarzschild geometry probes only $F_0$, while linear perturbations also probe magnetic and radiative curvature. The $M^{-(2N-2)}$ scaling of the shifts in $M\omega$ at fixed dimensionful coupling can help distinguish curvature orders. However, the cancellation for $K_6$ holds at every radius and mass, so combining Schwarzschild spectra at different masses cannot reveal this coupling. This limitation is intrinsic to the linearized dynamics, not a consequence of incomplete mode measurements.

The six-graviton amplitude distinguishes interactions that linear Schwarzschild spectroscopy cannot. Since the curvature expansion of $K_6$ about Schwarzschild contains cubic terms (Appendix A.2 of Ref.~\cite{companion}), nonlinear perturbations may also be sensitive to this interaction. Rotation offers another possibility, since perturbations about Kerr probe a different background Weyl curvature.

\begin{acknowledgments}
X.-H. F. is supported by the National Natural Science Foundation of China under Grants No.~11905157 and No.~11935009 and by the Tianjin University Self-Innovation Fund Extreme Basic Research Project under Grant No.~2025XJ21-0007. J. P. is supported by the National Natural Science Foundation of China under Grant No.~12405049.
\end{acknowledgments}

\begingroup
\setlength{\emergencystretch}{2em}

\endgroup

\onecolumngrid
\clearpage
\section*{Supplemental Material}
\begingroup
\linespread{1.12}\selectfont

We give the details of the large-multipole spectral proof and the finite-multipole calculation. The background solutions, full canonical potentials and proof of boundary equivalence are given in Ref.~\cite{companion}. All references are listed above.

\setlength{\abovedisplayskip}{6pt plus 2pt minus 1.5pt}
\setlength{\belowdisplayskip}{6pt plus 2pt minus 1.5pt}
\setlength{\abovedisplayshortskip}{4.5pt plus 2pt minus 1.5pt}
\setlength{\belowdisplayshortskip}{6pt plus 2pt minus 1.5pt}
\setcounter{secnumdepth}{2}
\setcounter{section}{0}
\setcounter{equation}{0}
\renewcommand{\thesection}{\Alph{section}}
\renewcommand{\thesubsection}{\thesection.\arabic{subsection}}
\renewcommand{\theHsection}{supp.\Alph{section}}
\numberwithin{equation}{section}
\renewcommand{\theHequation}{supp.\Alph{section}.\arabic{equation}}
\numberwithin{table}{section}
\renewcommand{\theHtable}{supp.\Alph{section}.\arabic{table}}
\setcounter{table}{0}
\section{Eikonal quasinormal modes}\label{app:large}
We use the potentials of Eqs.~(V.2)--(V.5) and Appendix E of Ref.~\cite{companion}, with explicit first-order curvature terms evaluated on Schwarzschild. Writing $D=f\partial_y$ and denoting $y$ derivatives by primes, we obtain the large-$l$ expansion
\begin{equation}
 \widehat V_\pm=J_lv_0+O(1)+\eps[J_l^2 W_\pm+J_l Z_\pm^{\rm eik}+O(1)].\label{sm:expansion}
\end{equation}
The squared GR potential in the radiative correction contributes $-\sigma_\pm f\tau/(8y^4)$ to $W_\pm$, where $\sigma_\pm=\pm1$ (odd: $-1$). The magnetic term contributes $-f\mu/(8y^4)$ to $W_-$, and the even potential from $F_0$ contributes $2f\phi'/y^3$, from the coefficient $C_1/J_l^2\to 2/y^3$ in Appendix E of Ref.~\cite{companion}. All other terms are at most linear in $J_l$. Substituting the homogeneous profiles gives Eq.~\eqref{eq:radial_leading}.

For $\mu=\tau=-16mpy^{-m}$, both leading $J_l^2$ corrections vanish identically. The GR odd-parity potential is $\widehat V_{-,0}=J_lv_0+v_1$, where $v_1=-6f/y^3$. Write $A=f+\eps A_1$. Extracting the $J_l$ coefficient of the odd potential gives
\begin{equation}
 Z_-^{\rm eik}=\frac{A_1}{y^2}+\frac{\tau v_0v_1}{4f}+(k_--2Da_{\tau,-})v_0-a_{\tau,-}Dv_0+\frac{f\tau}{4y^4}.
\end{equation}
The two auxiliary coefficients needed here are $a_{\tau,-}=-(2\tau/y-\tau')/8$ and $k_-=2z_-a_{\tau,-}+\tau z_-'/4$, where $z_-=f/y-f'/2$. At $y=3$ one has $Dv_0=z_-=0$. The homogeneous background solution in Eqs.~(III.11) and (III.12) of Ref.~\cite{companion}, with $A_1=b+2f\delta_1$, gives
\begin{equation}
 \left.\frac{A_1}{f}\right|_3=\frac{8(N-2)(3+1/N)}{27(N-1)}\frac{p}{3^m}.
\end{equation}
The remaining terms in $Z_-^{\rm eik}/v_0$ at the maximum combine to $-[(m+1)(m+2)/108]\tau(3)$.
Consequently
\begin{equation}
 \left.\frac{Z_-^{\rm eik}}{v_0}\right|_3=\frac{p}{27\,3^m}\left[\frac{8(N-2)(3+1/N)}{N-1}+4m(m+1)(m+2)\right].\label{nullresponse}
\end{equation}

For the barrier-top expansion at first order in the coupling, we introduce the semiclassical parameter $h=J_l^{-1/2}$ and $\mathfrak e=h^2\hat\omega^2$. The normalized Schr\"odinger operator $-h^2\partial_x^2+h^2\widehat V$ has leading-order potential barrier $v_0$ with a nondegenerate maximum at $y=3$. For the leading response set $\eta=\epsilon/h^2$; the normalized barrier is $v_0+\eta W_\pm+O(h^2)$ to first order in $\eta$. When both leading corrections vanish it is instead $v_0+\epsilon Z_-^{\rm eik}+O(h^2)$ to first order in $\epsilon$. In both cases $\partial_x=g\partial_y$ is regular near the maximum; its parameter dependence affects the barrier curvature, not the height at a stationary point.

Let $s$ be $\eta$ or $\epsilon$, and $v_s$ the corresponding leading-order barrier, maximal at $x_s$. For fixed overtone $j$, rescaling $x-x_s=\sqrt h\,\xi$ gives the inverted oscillator at order $h$. Barrier-top WKB theory~\cite{SchutzWill,BarrierTop} gives the formal series
\begin{equation*}
 \mathfrak e=v_s(x_s)-ih(j+\tfrac12)\sqrt{-2v_s''(x_s)}+O(h^2),
\end{equation*}
where derivatives are in $x$. The barrier height controls the leading oscillation frequency, while the curvature determines the damping. Since the order-$h$ term is imaginary for real $s$, the first-order real-frequency response in the WKB expansion is
\begin{equation*}
 \left.\partial_s\ln\Ree\hat\omega\right|_0=\frac{\left.\partial_s v_s(x_s)\right|_0}{2v_0(3)}+O(h^2).
\end{equation*}
The derivative of $x_s$ drops out by stationarity. For $s=\eta$ the left-hand side is $J_l^{-1}\partial_\epsilon\ln\Ree\hat\omega|_0$ and the numerator is $W_\pm(3)$; for $s=\epsilon$ with both leading corrections zero they are the unscaled derivative and $Z_-^{\rm eik}(3)$. Using $v_0(3)=1/27$ yields Eqs.~\eqref{eq:spectral_leading} and \eqref{null}.

\clearpage
\section{Numerical spectroscopy}\label{app:finite}
For the degree-six basis of Table~\ref{tab:qnm}, the dimensionless interactions are $(8192/27)X^3$, $(128/27)XP^2$ and $(512/81)\mathcal D_6$. We use the background and potentials of Secs.~III and V and Appendix E of Ref.~\cite{companion}, absorbing $\eps$ into the amplitudes. In particular, $B(y_H)=0$ and $\gamma_H=2M\kappa_H=\partial_y g(y_H)$ give $y_H=2-14\lambda_0/15$ and $\gamma_H=1/2+2\lambda_0/15$ through first order. With $z=y_H/y$, the ingoing and outgoing asymptotic factors are separated as
\begin{equation}
 \Psi=e^{i\hat\omega y_H/z}z^{-2i\hat\omega}(1-z)^{-i\hat\omega/\gamma_H}\chi(z).
\end{equation}
Let $q_z=z^2g/y_H$, $L_z=-y_H/z^2-2/z+1/[\gamma_H(1-z)]$ and $\mathcal K_z=q_z L_z$. The equation for the remaining radial amplitude is
\begin{equation}
 0=q_z\chi''+(q_z'+2i\hat\omega\mathcal K_z)\chi'+\left[\hat\omega^2\frac{1-\mathcal K_z^2}{q_z}+i\hat\omega\mathcal K_z' -\frac{\widehat V}{q_z}\right]\chi.
\end{equation}
Here primes denote $z$ derivatives. All coefficients are expanded at fixed $z$ to first order in the amplitudes, including the horizon displacement. At $z=0,1$, $q_z=0$ and $\mathcal K_z=-1,1$, respectively; the endpoint equations follow by taking the finite limits of the coefficients using their Taylor expansions. No additional Dirichlet conditions are imposed.

We discretize this equation, including its endpoint limits, by Chebyshev--Lobatto collocation \cite{Jansen}. The quadratic matrix eigenvalue problem $\mathsf Q(\hat\omega,\lambda)u_q=0$, where $\mathsf Q=\mathsf A+\hat\omega\mathsf B+\hat\omega^2\mathsf C$ is assembled from the collocation matrices, is solved by linearization, following the fundamental modes from their GR values. At a simple QNM eigenfrequency with right and left null vectors $u_q,v_q$, the first-order frequency shift is obtained from
\begin{equation}
 \partial_\lambda\hat\omega=-\frac{v_q^\dagger(\partial_\lambda\mathsf Q)u_q}{v_q^\dagger(\partial_{\hat\omega}\mathsf Q)u_q},\label{sm:eigenvalue_derivative}
\end{equation}
 evaluated at zero coupling for each response amplitude. Increasing the Chebyshev collocation degree from 40 to 56 changes the computed responses by less than $6\times10^{-11}$. Central differences with steps $\pm10^{-5}$ agree within $3\times10^{-7}$, and an independent Legendre--Lobatto discretization of degree 68 agrees with the Chebyshev results within $5.1\times10^{-11}$.

\endgroup
\end{document}